\documentclass[aps,prd,floats,floatfix,preprintnumbers,superscriptaddress,nofootinbib]{revtex4-2}

\usepackage{graphicx} 
\usepackage{epstopdf} 
\usepackage[normalem]{ulem}
\usepackage{amssymb}
\usepackage{amsmath}
\usepackage[hidelinks]{hyperref}
\usepackage{color}
\usepackage{xcolor,colortbl}
\usepackage[utf8]{inputenc}
\usepackage{slashed}
\usepackage{bm}

\newcommand{\be}{\begin{equation}}
\newcommand{\ee}{\end{equation}}
\newcommand{\bea}{\begin{eqnarray}}
\newcommand{\eea}{\end{eqnarray}}

\newcommand{\pd}{\partial}
\newcommand{\rmd}{\mathrm{d}}
\newcommand{\rme}{\mathrm{e}}
\newcommand{\rmi}{\mathrm{i}}
\newcommand{\calA}{\mathcal{A}}

\newcommand{\calP}{\mathcal{P}}

\newcommand{\calM}{\mathcal{M}}

\newcommand{\qonp}{\boldsymbol{q}_{1\perp}}
\newcommand{\qtwp}{\boldsymbol{q}_{2\perp}}

\newcommand{\lonp}{\boldsymbol{l}_{1\perp}}

\newcommand{\rp}{\boldsymbol{r}_{\perp}}
\newcommand{\kpi}{\boldsymbol{k}_{i\perp}}
\newcommand{\kpij}{\boldsymbol{k}_{i(j)\perp}}
\newcommand{\kpijj}{\boldsymbol{k}_{i(jj')\perp}}

\newcommand{\bp}{\boldsymbol{b}_{\perp}}

\newcommand{\lp}{\boldsymbol{l}_{\perp}}

\newcommand{\xp}{\boldsymbol{x}_{\perp}}
\newcommand{\yp}{\boldsymbol{y}_{\perp}}

\newcommand{\qp}{\boldsymbol{q}_{\perp}}
\newcommand{\qop}{\boldsymbol{q}_{1\perp}}
\newcommand{\qtp}{\boldsymbol{q}_{2\perp}}
\newcommand{\Sp}{\boldsymbol{S}_{\perp}}

\newcommand{\Pp}{\boldsymbol{P}_{\perp}}
\newcommand{\Kp}{\boldsymbol{K}_{\perp}}
\newcommand{\kp}{\boldsymbol{k}_{\perp}}

\newcommand{\delp}{\boldsymbol{\Delta}_{\perp}}
\newcommand{\dlp}{\boldsymbol{\delta}_{\perp}}

\newcommand{\epsp}{\boldsymbol{\epsilon}_{\perp}}
\newcommand{\kb}{\boldsymbol{k}}
\newcommand{\lambdab}{\bar{\lambda}}

\begin{document}
\date{\today}
\preprint{ZTF-EP-25-06}

\title{Off forward non-SCHC contributions to exclusive vector quarkonium production
from the ``spin dependent BFKL Pomeron''}

\author{Sanjin Beni\' c}
\email{sanjinb@phy.hr}
\affiliation{Department of Physics, Faculty of Science, University of Zagreb, Bijenička c. 32, 10000 Zagreb, Croatia}

\author{Adrian Dumitru}
\email{adrian.dumitru@baruch.cuny.edu}
\affiliation{Department of Natural Sciences, Baruch College, CUNY,
17 Lexington Avenue, New York, NY 10010, USA}
\affiliation{The Graduate School and University Center, The City University of New York, 365 Fifth Avenue, New York, NY 10016, USA}

\begin{abstract}
A novel contribution to off-forward, exclusive vector quarkonium
production, $\gamma^{(*)}+p \to V+p$, 
at high energy is derived which corresponds to a
$t$-channel exchange of a BFKL hard Pomeron, with a helicity flip of
the proton.   This ``spin-dependent BFKL Pomeron"
is required in a consistent expansion in powers of the momentum transfer {$ t \approx -\delp^2$}
beyond first order.
The spin-dependent Pomeron violates $s$-channel helicity
conservation (SCHC) at ${\cal O}(\Delta_\perp^2)$, and beyond.
{Expanding to leading twist only, it corresponds to GPD $E_g(x,t)$
for vanishing skewness.}
We derive explicit expressions for the eikonal BFKL
amplitudes, to all orders in dipole size times momentum transfer,
for all helicity configurations of the particles
in the initial and final states.
We also provide numerical estimates of the helicity flip
two gluon exchange amplitude at moderate $x$ from a light-cone quark model
of the proton.
The spin dependent BFKL Pomeron could, in principle, be discovered via 
double spin asymmetries in $e+p \to e+p+J/\psi$ with transversely polarized
proton and longitudinally polarized electron in the initial state.
\end{abstract}

\maketitle

\section{Introduction}
The eikonal, forward BFKL Pomeron or unintegrated gluon
distribution~\cite{Lipatov:1976zz,Kuraev:1977fs,Balitsky:1978ic,Mueller:1993rr}
corresponds to the matrix element of a two-gluon
$t$-channel exchange operator between incoming and outgoing
light-cone proton states with the same helicity {\color{black} $\lambda_p = \pm 1/2$}. 
Indeed, in the
eikonal (high-energy) and $t\to0$ limits a helicity flip of the
proton is not possible; the same applies to the transition
of the photon to the vector meson state. This leads to so-called $s$-channel
helicity conservation
(SCHC) in eikonal exclusive $J/\psi$ or $\psi(2S)$ production when $t \to 0$~\cite{Ivanov:2004ax}.
Specifically, in photoproduction or ultraperipheral
proton-proton or nucleus-proton collisions, the vector meson (VM) is
transversely polarized, and in fact its polarization $\bar\lambda$
is {\em equal} to the polarization $\lambda=\pm1$ of the incoming photon.
\\

In off-forward production, $-t > 0$, non-SCHC corrections appear.
The first kind preserves still the helicity of the proton, and is
thus associated with the usual spin independent non-forward BFKL
Pomeron~\cite{Ivanov:2004ax,Nemchik:1994fp,Kuraev:1998ht} involving orbital angular momentum transfer to
the $c-\bar{c}$ pair which scatters off the proton. Thus, the
polarization of the $J/\psi$ may now be different from that of
the photon, $\bar\lambda \ne \lambda$. In the limit of small $|t|
\approx \delp^2$, non-SCHC contributions to the
cross section in DIS (photoproduction) begin at linear (quadratic) order in the VM transverse momentum
$\Delta_\perp$.

Here, we are concerned mainly with a second kind of non-SCHC corrections
associated with a helicity flip of the proton. These are related to
the so-called ``spin-dependent Pomeron" (SDP)~\cite{Boussarie:2019vmk,Hagiwara:2020mqb,Hatta:2022bxn,Agrawal:2023mzm}.
Hard processes involve the partonic
substructure of the proton, and orbital angular momentum may be
transfered to the partons even in the eikonal limit, thereby
allowing a helicity flip of the proton\footnote{The SDP is associated with the $n=0$ BFKL eigenfunction
with positive energy intercept.}. The SDP
is parameterized
in terms of two distinct dipole-proton scattering amplitudes, associated with $(\rp\cdot\delp)(\epsp^\Lambda\cdot\rp)$ 
and $(\epsp^\Lambda\cdot\delp)$ correlations, where $\rp$ is the dipole vector and $\epsp^\Lambda = -
(\Lambda,\rmi)/\sqrt{2}$ the helicity vector; {\color{black} hereafter we conventionally normalize the} proton helicity defining $\Lambda = {\color{black} 2\lambda_p} = \pm 1$. 
Therefore, the SDPs contribute
at leading non-SCHC power ${\cal O}(\Delta_\perp^2)$ to the cross
section for exclusive vector quarkonium photoproduction (at small
momentum transfer). 

The ZEUS and H1 collaborations at HERA have confirmed that the cross section ratio for
longitudinal vs.\ transverse $J/\psi$ polarization in photoproduction
on unpolarized protons is small~\cite{ZEUS:2002wfj,H1:2005dtp}. Similarly,
these experiments found that
the amplitude for $J/\psi$ helicity opposite to photon helicity is much smaller
than the SCHC amplitude. However, these data do not constrain the
contribution associated with a helicity flip of the proton.

A polarisation measurement of coherently photoproduced $J/\psi$ in ultra-peripheral Pb-Pb collisions at
the LHC has been reported by the ALICE collaboration~\cite{ALICE:2023svb,Massacrier:2024fgx}.
These data again are found to be consistent with SCHC. In a model where the photon
scatters off a ``Pomeron patch" in the nucleus~\cite{Brandenburg:2024ksp}
there could be a contribution though which involves a helicity flip within that patch.

The STAR collaboration at RHIC has constrained the presence of a hadronic spin flip due to 
{\em soft} Pomeron exchange in polarized proton–proton elastic scattering at very low $|t|
\le 0.035$~GeV$^2$~\cite{STAR:2012fiw}.
This process lacks a hard scale though and so it does not relate in a computable way to the
gluon (ladder) exchanges of perturbative QCD. Also, as the numerical estimates below will show,
a significant SDP amplitude may require higher momentum transfer, $\Delta_\perp \sim 0.5$~GeV or greater.
\\

{In Sec.~\ref{sec:amps} we introduce the general BFKL eikonal amplitudes for exclusive VM production. The following Sec.~\ref{sec:sip} is focused on their helicity decomposition and in Sec.~\ref{sec:sdp} we consider the proton helicity flip case. In Sec.~\ref{sec:num} we perform a model computation of the SDPs for moderately small $x$.
Sec.~\ref{sec:conc} concludes the paper with a summary and discussion.}

\section{The $\gamma^{(*)}+p \to V+p$ amplitudes}
\label{sec:amps}
The general formula for the eikonal amplitude of exclusive VM production $\gamma(q,\lambda) p(P,\Lambda) \to V(\Delta,\lambdab) p'(P',\Lambda')$ is
\be
\calM_{\lambda\Lambda;\lambdab\Lambda'} = 2 N_c \int \rmd^2 \rp \calP_{\Lambda\Lambda'}(\rp,\delp) \calA_{\lambda \lambdab}(\rp,\delp)\,,
\label{eq:amp}
\ee
in terms of which the $S$-matrix reads $(2\pi)\delta(q^ - - \Delta^-) q^- \calM_{\lambda\Lambda;\lambdab\Lambda'}$, after subtracting the no-scattering contribution. Our notation mostly follows ref.~\cite{Mantysaari:2020lhf}, and our amplitudes below without proton helicity flip, $\Lambda' = \Lambda$, agree with ref.~\cite{Mantysaari:2020lhf}
except for different signs of the $\gamma_L \to V_L$ and $\gamma_\lambda \to V_{-\lambda}$ amplitudes.
In eq.~(\ref{eq:amp}) the helicities of the incoming and outgoing proton, as well as of the photon and the VM are indicated
explicitly. $\calA_{\lambda\lambdab}(\rp,\delp)$ is the reduced amplitude representing the $\gamma-V$ wave function overlap. For a transverse photon with $\lambda = \pm 1$, the helicity dependence of the $\gamma-V$ overlaps reads
\be
\begin{split}
& \calA_{\lambda = \pm 1,\lambdab = 0}(\rp,\delp) = e q_c \lambda \rme^{\rmi \lambda \phi_r} \int_z \rme^{-\rmi \dlp\cdot\rp} \calA_{TL}(r_\perp,z)\,,\\
&\calA_{\lambda = \pm 1,\lambdab = \pm 1}(\rp,\delp) = e q_c \int_z \rme^{-\rmi \dlp\cdot \rp}\left[\delta_{\lambda\lambdab}\calA^{nf}_{TT}(r_\perp,z) + \delta_{\lambda,-\lambdab}\rme^{2\rmi\lambda\phi_r}\calA_{TT}^f(r_\perp,z)\right]\,.
\end{split}
\label{eq:overlaps}
\ee
Here, $\dlp = (z-\bar z)\delp/2$ where $z$ and $\bar z = 1-z$ denote the momentum fractions of the quark and anti-quark
in the VM, respectively. We also have $\int_z \equiv \int_0^1 \frac{\rmd z}{4\pi}$. In the non-relativistic heavy-quark limit, the $q\bar{q}$ LC momentum imbalance is $z-\bar z \to 0$.
The functions $\calA_{TL}(r_\perp,z)$ and $\calA_{TT}^{nf,f}(r_\perp,z)$ are collected in Appendix \ref{sec:A_i(r,z)}, together with the $\gamma_L-V$ overlaps.

In eq.~(\ref{eq:amp}), $\calP_{\Lambda\Lambda'}(\rp,\delp)$ represents BFKL Pomeron exchange, defined as
\be
\calP_{\Lambda\Lambda'}(\rp,\delp) = \int\rmd^2 \bp \rme^{-\rmi \delp\cdot\bp}\, \left\{
1 - \frac{1}{2 N_c}\left[\frac{\langle P' \Lambda'|{\rm tr}\left[V(\xp)V^\dag(\yp) + V(\yp)V^\dag(\xp)\right]|P\Lambda\rangle}{\langle P\Lambda|P\Lambda\rangle}\right]\right\}\,,
\label{eq:pom1}
\ee
with $\xp = \bp + \rp/2$, $\yp = \bp - \rp/2$. The eikonal Wilson lines $V(\xp) = P \exp\left[-\rmi g \int_{-\infty}^\infty \rmd x^- A^+(x^-,\xp)\right]$, with $A^+(x^-,\xp) = t_a A_a^+(x^-,\xp)$ the gluon field in covariant gauge, are used to construct the dipole amplitude ${\rm tr}\left[V(\xp)V^\dag(\yp)\right]/N_c$. In \eqref{eq:pom1}  we pick up its $C$-even part, as dictated by the symmetries of the photon and the VM wave functions. The amplitude $\calP_{\Lambda \Lambda'}(\rp,\delp)$ can be parametrized in the most general way as \cite{Boussarie:2019vmk}
\be
\calP_{\Lambda\Lambda'}(\rp,\delp) = \delta_{\Lambda,\Lambda'}\calP(\rp,\delp) + \cos(\phi_{r\Delta})\Lambda\rme^{\rmi \Lambda\phi_r}\delta_{\Lambda,-\Lambda'}\calP_S(\rp,\delp) +\Lambda\rme^{\rmi \Lambda\phi_\Delta}\delta_{\Lambda,-\Lambda'}\calP^\perp_S(\rp,\delp)\,,
\label{eq:pom2}
\ee
where $\phi_{ab} = \phi_a - \phi_b$ and $\calP(\rp,\delp)$, $\calP_S(\rp,\delp)$ and $\calP^\perp_S(\rp,\delp)$ are three in principle different non-perturbative real scalar functions. In App.~\ref{sec:gtmds} we briefly recall the connection to the GTMDs. $\calP(\rp,\delp)$ is the ``spin-independent Pomeron" (SIP) that is usually employed in calculations of $J/\psi$ production in high-energy photon-proton or photon-nucleus 
scattering~\cite{Nemchik:1994fp,Kuraev:1998ht,Ivanov:2004ax,Kowalski:2006hc,Armesto:2014sma,Mantysaari:2020lhf,Mantysaari:2023xcu,Boer:2023mip,Penttala:2024hvp}. The SDPs $\calP_S(\rp,\delp)$ and $\calP^\perp_S(\rp,\delp)$, associated with a helicity-flip of the proton, have so far not been
identified in the context of eikonal, exclusive VM production. In the collinear limit, $\calP$ is related to the gluon GPD $H_g(x,t)$ at small-$x$. Likewise, the combination $\calP_S + 2 \calP^\perp_S$ is related to the GPD $E_g(x,t)$ \cite{Hagiwara:2020mqb,Hatta:2022bxn}. The helicity-flip part should be linear in $\epsp^\Lambda$ which explains the angular structure in \eqref{eq:pom2} after taking into account that $\calP_{\Lambda\Lambda'}(\rp,\delp)$ is even in $\rp \to - \rp$. In this parametrization $\calP_S \sim \Delta_\perp$ and $\calP_S^\perp \sim \Delta_\perp$ for small $\Delta_\perp$. Each of the scalar functions depend on the variables $\rp^2$, $\delp^2$, $\rp\cdot\delp$ which can be revealed through a Fourier expansion, for example
\be
\calP(\rp,\delp) = \calP_0(r_\perp,\Delta_\perp) + 2\calP_\epsilon(r_\perp,\Delta_\perp)\cos(2\phi_{r\Delta})
 + \dots\,,
\label{eq:four}
\ee
and similarly for $\calP_S(\rp,\delp)$ and $\calP^\perp_S(\rp,\delp)$.
The appearance of the ``elliptic" Pomeron $\calP_\epsilon(r_\perp,\Delta_\perp)$ in DVCS at small $x$ was first pointed out in ref.~\cite{Hatta:2017cte} and in connection to the gluon GPD $E_{gT}(x,t)$. Its relevance to exclusive VM production was mentioned in ref.~\cite{Mantysaari:2020lhf},
we add below insight into the specific helicity channels and the power counting in $\Delta_\perp$.
The azimuthal angular dependence of the dipole scattering amplitude on a proton, specifically, has been
discussed in ref.~\cite{Dumitru:2021tvw}. In the limit of small $\Delta_\perp$,
$\calP_0(r_\perp,\Delta_\perp) \sim \Delta_\perp^0$ while $\calP_\epsilon(r_\perp,\Delta_\perp) \sim \Delta_\perp^2$;
higher Fourier harmonics are proportional to higher powers of $\Delta_\perp$ and are expected to have very small
amplitudes, so we will not consider them further.

\section{Spin-independent Pomeron exchange}
\label{sec:sip}
We first list the amplitudes for Pomeron exchange without helicity
flip of the proton and discuss their scaling with $\Delta_\perp$ in the
limit of small momentum transfer\footnote{That is, the scaling with $\Delta_\perp$
of the leading twist contribution to the respective amplitude. However, in the BFKL approach
one does not expand in powers of $r_\perp \Delta_\perp$, and high transverse momentum transfer is allowed.}.
Even in the absence of a helicity flip of
the proton we obtain new contributions to the cross section at
${\cal O}(\Delta_\perp^2)$ as compared to the classic work
by Nikolaev et al.~\cite{Nemchik:1994fp,Kuraev:1998ht,Ivanov:2004ax}, which are due to 
interference of the SCHC $\gamma-V$ amplitude with the helicity flip $\gamma-V$ amplitude for the elliptic Pomeron, or gluon distribution.

We begin with the amplitudes for a transverse photon ($\lambda = \pm 1$) which survive in the $Q^2\to0$ photoproduction limit.
The amplitude for a transverse VM can be decomposed into helicity non-flip ($\lambda = \lambdab$) and helicity flip ($\lambda = -\lambdab$) pieces.
From the first term in the Fourier series \eqref{eq:four}, i.e.\ the isotropic Pomeron,
the helicity non-flip SCHC amplitude is
\be
\calM_{\lambda\Lambda;\lambda\Lambda}= 4\pi N_c e q_c\int_z \int r_\perp \rmd r_\perp
  \calP_0(r_\perp,\Delta_\perp)
  \mathcal A^{nf}_{TT}(r_\perp,z)\, J_0(r_\perp \delta_\perp)
 \sim \Delta_\perp^0 \,,
  \label{eq:M_TT_nf0}
\ee
Recall that ${\delta}_\perp={|z-\bar{z}|}\Delta_\perp/2$.
For small momentum transfer this amplitude is independent of $\Delta_\perp$.
{For completeness we also list the non-flip amplitude for the elliptic Pomeron
which starts out at order ${\cal O}(\Delta_\perp^4)$:}
\be
\calM^\epsilon_{\lambda\Lambda;\lambda\Lambda}= -8\pi N_c e q_c \int_z \int r_\perp \rmd r_\perp
  \calP_\epsilon(r_\perp,\Delta_\perp)
  \mathcal A^{nf}_{TT}(r_\perp,z)\, J_2(r_\perp \delta_\perp)
 \sim \Delta_\perp^4 \,.
  \label{eq:M_TT_nf2}
\ee
On the other hand, the $\gamma-V$ helicity flip amplitude is
\be
  \calM_{\lambda\Lambda;-\lambda\Lambda}= 4\pi N_c e q_c \rme^{2\rmi \lambda\phi_\Delta}
  \int_z\int r_\perp \rmd r_\perp
  \calP_0(r_\perp,\Delta_\perp)
  \mathcal A^{f}_{TT}(r_\perp,z) J_2(r_\perp \delta_\perp)
  \sim \Delta_\perp^2  \,.
  \label{eq:M_TT_f0}
\ee
This amplitude scales as
$\Delta_\perp^2$ due to the transfer of two units of orbital angular
momentum {to compensate for the helicity flip}. Each unit of orbital angular momentum comes with
one power of $\Delta_\perp$, one power of the dipole size $r_\perp$,
and finally one power of the LC momentum imbalance  $|z-\bar z|$.
Hence, we expect that the above non-SCHC contribution is
smaller for $\Upsilon$ than for $J/\psi$ but greater for $\psi(2S)$ than $J/\psi$ mesons
as the transverse size of the $\psi(2S)$ and the width of its light-cone
distribution amplitude~\cite{Hwang:2008qi} about $z=1/2$ are greater than those of the $J/\psi$.

For low momentum transfer the second term in \eqref{eq:four}, i.e.\ the elliptic Pomeron, scales as $\sim\Delta_\perp^4$ for $\lambdab = \lambda$, but as $\sim\Delta_\perp^2$ for $\lambdab = - \lambda$,
\be
\begin{split}
& \calM^\epsilon_{\lambda\Lambda;-\lambda\Lambda} = 4\pi e q_c N_c {\color{black}\rme^{2\rmi\lambda \phi_\Delta}} \int_z \int r_\perp \rmd r_\perp \calP_\epsilon(r_\perp,\Delta_\perp) \calA^{f}_{TT}(r_\perp,z)\left[J_4(r_\perp \delta_\perp) + J_0(r_\perp \delta_\perp)\right] \sim \Delta_\perp^2\,,
  \label{eq:M_TT_f2}
\end{split}
\ee
without suppression by factors of $|z-\bar{z}|$.

Coming to the longitudinal VM, the first term in \eqref{eq:four} leads to
\be
\calM_{\lambda\Lambda;0\Lambda} = -4\pi \rmi e q_c N_c \lambda\rme^{\rmi \lambda\phi_\Delta} \int_z {\rm sign}(z-\bar{z})\int r_\perp \rmd r_\perp \calP_0(r_\perp,\Delta_\perp) \calA_{TL}(r_\perp,z) \, J_1(r_\perp \delta_\perp)
\sim \Delta_\perp^1 \,.
\label{eq:M_TL_0}
\ee
This amplitude is proportional to a single power of $\Delta_\perp$ and a single power of $r_\perp$ (at small
$\Delta_\perp$) but is quadratic in the LC momentum imbalance $|z-\bar z|$ since $\calA_{TL}(r_\perp,z) \sim
z-\bar z$, see eq.~\eqref{eq:ATL2}. The elliptic Pomeron contributes at order $\Delta_\perp^3$:
\be
\calM^\epsilon_{\lambda\Lambda;0\Lambda} = 
-4\pi \rmi e q_c N_c \lambda {\color{black} \rme^{\rmi\lambda\phi_\Delta}}
\int_z {\rm sign}(z-\bar{z})\int r_\perp \rmd r_\perp 
\calP_\epsilon(r_\perp,\Delta_\perp) \calA_{TL}(r_\perp,z) \, 
\left[J_1(r_\perp \delta_\perp) - J_3(r_\perp \delta_\perp)\right]
\sim \Delta_\perp^3 \,.
\ee
%
Therefore, a test of SCHC violation via the cross section ratio for longitudinal vs. transverse VM production $\sigma_L / \sigma_T$ involves
a suppression factor
\be
\frac{\sigma_L}{\sigma_T}
~\sim   ~ \left< (z-\bar z)^4 \, r^2_\perp\right>_V \, \Delta_\perp^2~.
\label{eq:sigmaL/sigmaT}
\ee
On the other hand, the interference of the amplitude for
a $\gamma-V$ helicity flip by two units, $\calM_{\lambda\Lambda;-\lambda\Lambda}$ in \eqref{eq:M_TT_f0}, with the non-flip amplitude
$\calM_{\lambda\Lambda;\lambda\Lambda}$ in \eqref{eq:M_TT_nf0}, is suppressed by two powers of $z-\bar z$:
\be
\frac{\calM_{\lambda\Lambda;-\lambda\Lambda} \calM^*_{\lambda\Lambda;\lambda\Lambda} + \mathrm{c.c.}}
{\left| \calM_{\lambda\Lambda;\lambda\Lambda}\right|^2}
~\sim   ~ \left< (z-\bar z)^2\, r^2_\perp\right>_V \, \Delta_\perp^2~.
\ee
There is also an interference of $\calM^\epsilon_{\lambda\Lambda;-\lambda\Lambda}$ and $\calM_{\lambda\Lambda;\lambda\Lambda}$  which is not
suppressed by powers of $z-\bar z$ at small $\Delta_\perp$. However,
it involves the ratio of the elliptic to the isotropic BFKL unintegrated gluon distributions,
see below.

We now list the amplitudes for a longitudinal virtual photon ($\lambda = 0$). These are
\be
\begin{split}
\calM_{0\Lambda;0\Lambda} &= 4\pi e q_c N_c \int_z \int r_\perp \rmd r_\perp \calP_0(r_\perp,\Delta_\perp)\calA_{LL}(r_\perp,z) J_0(r_\perp\delta_\perp)\sim \Delta_\perp^0\,,\\
\calM^\epsilon_{0\Lambda;0\Lambda} &= - 8\pi e q_c N_c \int_z \int r_\perp \rmd r_\perp \calP_\epsilon(r_\perp,\Delta_\perp)\calA_{LL}(r_\perp,z) J_2(r_\perp\delta_\perp)\sim \Delta_\perp^4\,, \\ 
\end{split}
\ee
where $\calA_{LL}(r_\perp,z)$ is given in the first line of eq.~\eqref{eq:Along}.
The first term, together with \eqref{eq:M_TT_nf0}, represent the usual SCHC $\gamma-V$ amplitudes associated with the non-flip $T\to T$ and the $L\to L$ transitions. 
For the $L\to T$ transition we have
\be
\begin{split}
\calM_{0\Lambda;\lambdab\Lambda} &= -4\pi \rmi e q_c N_c \lambdab\rme^{-\rmi \lambdab\phi_\Delta} \int_z \int r_\perp \rmd r_\perp \calP_0(r_\perp,\Delta_\perp) \calA_{LT}(r_\perp,z) \, {\rm sign}(z-\bar{z})J_1(r_\perp \delta_\perp)
\sim \Delta_\perp^1\,,\\
\calM^\epsilon_{0\Lambda;\lambdab\Lambda} &= 4\pi \rmi e q_c N_c \rme^{-\rmi \lambdab\phi_\Delta} \int_z \int r_\perp \rmd r_\perp \calP_\epsilon(r_\perp,\Delta_\perp) \calA_{LT}(r_\perp,z) 
\left(J_{2-\lambdab}(r_\perp \delta_\perp) - J_{2+\lambdab}(r_\perp \delta_\perp)\right)
\sim \Delta_\perp^3\,,
\label{eq:M_LT_0}
\end{split}
\ee
with $\calA_{LT}(r_\perp,z)$ in the second line of eq.~\eqref{eq:Along}.

Let us summarize all contributions to the VM cross section due
to the SIP exchange, organized
by powers of $\Delta_\perp$. At leading power $\Delta_\perp^0$, we have the squares of
the SCHC amplitudes $\calM_{\lambda\Lambda;\lambda\Lambda}$ and $\calM_{0\Lambda;0\Lambda}$, which are the only
amplitudes that survive in the forward limit. 
At first power in momentum transfer $\Delta_\perp$, and for non-zero $Q^2$, there is a contribution due to
interference of $\calM_{0\Lambda;\lambdab\Lambda}$ and $\calM_{\lambda\Lambda;\lambda\Lambda}$ (for
$\lambdab=\lambda$) as
well as of $\calM_{0\Lambda;0\Lambda}$ and $\calM_{\lambda\Lambda;0\Lambda}$.
At order $\Delta_\perp^2$, and in the $Q^2\to0$ photoproduction limit, we have 
i) the interference of the SCHC amplitude
$\calM_{\lambda\Lambda;\lambda\Lambda}$ with the sum of the helicity flip
amplitudes $\calM_{\lambda\Lambda;-\lambda\Lambda}+\calM^\epsilon_{\lambda\Lambda;-\lambda\Lambda}$,
and ii) the square of the $T\to L$ amplitude
$\calM_{\lambda\Lambda;0\Lambda}$ for the isotropic gluon distribution.
For high momentum transfer of order of the mass $m_c$ of the heavy quark one needs
to sum all of the above amplitudes, of course.
This is the regime we are focusing on.

\section{Spin-dependent Pomeron exchange, proton helicity flip}
\label{sec:sdp}
The leading angular dependence of the SDPs $\calP_S(\rp,\delp)$ and $\calP^\perp_S(\rp,\delp)$ is given in \eqref{eq:pom2}.
Retaining only the first Fourier harmonics, $\calP_S(\rp,\delp) \approx \calP_{S0}(r_\perp,\Delta_\perp)$ and $\calP^\perp_S(\rp,\delp) \approx \calP^\perp_{S0}(r_\perp,\Delta_\perp)$,
the resulting amplitudes for a transverse photon, and their
scaling with $\Delta_\perp$ for $\Delta_\perp\to0$, are
\be
\begin{split}
\calM_{\lambda \Lambda; 0, -\Lambda} &=
{2}\pi\rmi  N_c e q_c \rme^{\rmi(\lambda+\Lambda)\phi_\Delta} \int_z \mathrm{sgn}(z-\bar{z}) \int
r_\perp \rmd r_\perp \,\mathcal A_{TL}(r_\perp,z) \calP_{S0}(r_\perp,\Delta_\perp) \left[
  J_{\lambda+\Lambda+1}(r_\perp\delta_\perp) - J_{\lambda+\Lambda-1}(r_\perp\delta_\perp) \right]  \sim \Delta_\perp^2~,\\
  \calM_{\lambda \Lambda, \lambda, -\Lambda} &=
-2\pi  N_c e q_c {\Lambda}\rme^{\rmi\Lambda\phi_\Delta} \int_z\int
r_\perp \rmd r_\perp \mathcal A_{TT}^{nf}(r_\perp,z) \calP_{S0}(r_\perp,\Delta_\perp) {\left[
  J_2(r_\perp\delta_\perp) - J_0(r_\perp\delta_\perp) \right]}  \sim \Delta_\perp^1~,\\
\calM_{\lambda \Lambda; -\lambda, -\Lambda} &=
2\pi  N_c e q_c \rme^{\rmi(2\lambda+\Lambda)\phi_\Delta} \int_z \int
r_\perp \rmd r_\perp \mathcal A_{TT}^{f}(r_\perp,z) \calP_{S0}(r_\perp,\Delta_\perp) \left[
  J_{2\lambda+\Lambda+1}(r_\perp\delta_\perp) - J_{2\lambda+\Lambda-1}(r_\perp\delta_\perp) \right]  \sim \Delta_\perp^{2+\Lambda\lambda}~,
\end{split}
\label{eq:Msdp}
\ee
%
%
and
\be
\begin{split}
\calM_{\lambda \Lambda; 0, -\Lambda} &=
- 4\pi \rmi N_c e q_c \Lambda \rme^{\rmi(\Lambda+\lambda)\phi_\Delta} \int_z \mathrm{sgn}(z-\bar{z}) \int
r_\perp \rmd r_\perp \, \mathcal A_{TL}(r_\perp,z) \calP_{S0}^\perp(r_\perp,\Delta_\perp)
  J_{\lambda}(r_\perp\delta_\perp)  \sim \Delta_\perp^2~,\\
\calM_{\lambda \Lambda; \lambda, -\Lambda} &=
4\pi  N_c e q_c \Lambda \rme^{\rmi\Lambda\phi_\Delta}  \int_z \int
r_\perp \rmd r_\perp \mathcal A_{TT}^{nf}(r_\perp,z) \calP^\perp_{S0}(r_\perp,\Delta_\perp)
  J_0(r_\perp\delta_\perp) \sim \Delta_\perp^1~,\\
\calM_{\lambda \Lambda; -\lambda, -\Lambda} &=
- 4\pi  N_c e q_c  \Lambda \rme^{\rmi(2\lambda+\Lambda)\phi_\Delta}  \int_z \int
r_\perp \rmd r_\perp \mathcal A_{TT}^{f}(r_\perp,z) \calP^\perp_{S0}(r_\perp,\Delta)
  J_{2}(r_\perp\delta_\perp)  \sim \Delta_\perp^3~.
\end{split}
\label{eq:Msdpperp}
\ee
Therefore, SDP exchanges associated with a helicity flip of the
proton contribute to the VM photoproduction cross section at ${\cal O}(\Delta_\perp^2)$
through all the ${\cal O}(\Delta_\perp)$ amplitudes from above, incl.\ their
interference.  These are the two $\calM_{\lambda \Lambda;
  \lambda, -\Lambda}(\delp)$ without a helicity flip from photon to VM, as well as the double helicity flip amplitude
$\calM_{\lambda \Lambda; -\lambda,-\Lambda}(\delp)$ with the first type of
SDP exchange, and for opposite helicities of the incoming photon and
proton. Note that at leading power of $r_\perp \delta_\perp$, i.e.\ at leading twist, 
each amplitude of eq.~(\ref{eq:Msdp})
combines with the corresponding amplitude of~(\ref{eq:Msdpperp}) to $E_g \sim \calP_{S0} + 2\calP_{S0}^\perp$, in agreement with ref.~\cite{Koempel:2011rc}.
We also point out that, interestingly, none of these amplitudes
involves factors of $z-\bar z$. Hence, for small momentum transfer $\Delta_\perp$,
the ratio of proton helicity flip to non-flip cross sections scales as
\be
\frac{\sigma_{\Lambda'=-\Lambda}}{\sigma_{\Lambda'=\Lambda}} \sim \left< r_\perp^2\right>_V\, \Delta_\perp^2~,
\label{eq:sigfnf}
\ee
Eq.~\eqref{eq:sigfnf} lacks the large $(z-\bar z)^4$ suppression factor of eq.~(\ref{eq:sigmaL/sigmaT}). 
On the other hand,
the coefficient involves the squared ratio of the helicity flip Pomeron to the conventional BFKL Pomeron
which at present is unknown. Below we present a first estimate of the
eikonal, helicity flip two-gluon exchange amplitude for moderately small $x$ from a
non-perturbative LC quark model of the proton.

For a longitudinal virtual photon,
\be
\begin{split}
\calM_{0\Lambda;0,-\Lambda} & = -2\pi e q_c N_c \Lambda\rme^{\rmi \Lambda\phi_\Delta} \int_z \int r_\perp \rmd r_\perp \calP_{S0}(r_\perp,\Delta_\perp) \calA_{LL}(r_\perp,z) \left[J_2(r_\perp \delta_\perp) - J_0(r_\perp\delta_\perp)\right]
\sim \Delta_\perp^1 \,,\\
\calM_{0 \Lambda; \lambdab, -\Lambda} & =
-2\pi\rmi  N_c e q_c \rme^{\rmi(-\lambdab+\Lambda)\phi_\Delta} \int_z \mathrm{sgn}(z-\bar{z}) \int
r_\perp \rmd r_\perp \,\mathcal A_{LT}(r_\perp,z) \calP_{S0}(r_\perp,\Delta_\perp) \\
&\times\left[
  J_{-\lambdab+\Lambda+1}(r_\perp\delta_\perp) - J_{-\lambdab+\Lambda-1}(r_\perp\delta_\perp) \right]  \sim \Delta_\perp^2\,,\\
\label{eq:M_LT_2}
\end{split}
\ee
and
\be
\begin{split}
&\calM_{0\Lambda;0,-\Lambda} = 4\pi N_c e q_c \Lambda\rme^{\rmi \Lambda\phi_\Delta} \int_z \int r_\perp \rmd r_\perp \calP^\perp_{S0}(r_\perp,\Delta_\perp) \calA_{LL}(r_\perp,z) J_0(r_\perp \delta_\perp)
\sim \Delta_\perp^1\,,\\
&\calM_{0 \Lambda; \lambdab, -\Lambda} =
- 4\pi \rmi N_c e q_c \Lambda \rme^{\rmi(-\lambdab+\Lambda)\phi_\Delta} \int_z \mathrm{sgn}(z-\bar{z}) \int
r_\perp \rmd r_\perp \, \mathcal \calP_{S0}^\perp(r_\perp,\Delta_\perp)A_{LT}(r_\perp,z)
  J_{\lambdab}(r_\perp\delta_\perp)  \sim \Delta_\perp^2~.
\end{split}
\ee
Hence, at non-zero photon virtuality we obtain an additional contribution to the cross section at ${\cal O}(\Delta_\perp^2)$.

Isolating the contribution of the SDP requires access to the spin flip amplitude. We comment on the suggestion of Refs.~\cite{Koempel:2011rc,Massacrier:2024fgx} to use target single spin asymmetries via the cross section difference $\Delta\sigma(\Sp) = \sigma(\Sp) - \sigma(-\Sp)$, where the initial proton is transversely polarized\footnote{An obvious alternative would be to detect the polarization of the recoil proton, but to our knowledge this is not part of the EIC polarimetry program, though there are proposals for such measurements at JLab \cite{BessidskaiaBylund:2022qgg}.}  with spin $\Sp$. Since the proton's spin enters the cross section alongside a factor of `i', $\Delta\sigma(\Sp)$ ends up probing the {\it imaginary} part of the interference between the proton helicity flip and the non-flip amplitudes
\be
\Delta\sigma(\Sp) \sim {\rm Im}[\calM_{nf}(\gamma \to V) \calM_f^*(\gamma\to V)]\,,
\ee
However, in the high energy limit employed here, the corresponding amplitudes $\calM_{nf}$ and $\calM_f$ are real, and so single spin asymmetries vanish. The real part of the interference can be captured through the {\it double} spin asymmetry $\Delta\sigma(\lambda_e = +1,\Sp)-\Delta\sigma(\lambda_e = -1,\Sp)$ where the incoming electron beam has longitudinal polarization $\lambda_e$. The presence of an additional spin supplies another factor of `$\rmi$' and so:
\be
\Delta\sigma(\lambda_e = +1,\Sp)-\Delta\sigma(\lambda_e = -1,\Sp)\sim{\rm Re}\left[\calM_{nf}(\gamma_L \to V)\calM^*_{f}(\gamma_T \to V) - \calM_{nf}(\gamma_T \to V)\calM^*_{f}(\gamma_L \to V)\right]\,.
\ee
A non-vanishing asymmetry is now realized thanks to the interference between longitudinal and transverse photons.

\section{Estimate of the spin dependent Pomeron at moderately small $x$}
\label{sec:num}
We now provide a numerical estimate of the magnitude of the spin dependent Pomeron at moderately small $x$
from a light-front constituent quark model for the proton. Of course, such models lack a deeper
theoretical justification and should be viewed as empirically motivated parameterizations of
the structure of the proton at moderate and large $x$ (and low resolution scales). In particular,
some recent ideas about the fundamental, non-perturbative origin of the spin of the proton
refer to the role of anomalous interactions and 
topology~\cite{Dorokhov:1993ym,Kochelev:2003cp,Kochelev:2015pqd,Zhang:2017zpi,Shuryak:2021fsu,Shuryak:2022thi,Miesch:2023hjt,Tarasov:2021yll,Tarasov:2025mvn}, 
with the difference in the number of left- vs.\ right-handed quarks
given by the topological charge density in the proton. The light-front constituent quark model we employ
assumes that the helicity wave function of the proton emerges simply by Melosh transformation of
non-relativistic Pauli spinors for massive quarks to the light-front.

Empirically motivated light-front quark models have been used
extensively in the literature to compute Dirac and Pauli form factors and anomalous
magnetic moments~\cite{Schlumpf:1992vq,Brodsky:1994fz,Miller:2007uy} of proton and neutron,
transverse momentum dependent (TMD) parton distributions~\cite{Ji:2002xn,Pasquini:2008ax,Pasquini:2010af},
generalized parton distributions (GPDs)~\cite{Ji:2002xn},
gravitational form factors~\cite{Chakrabarti:2020kdc,Choudhary:2022den,Nair:2024fit},
spin independent two-gluon (Pomeron) and three-gluon (Odderon) 
amplitudes~\cite{Dumitru:2018vpr,Dumitru:2019qec,Dumitru:2020fdh,Dumitru:2021tvw},
quark Wigner distributions~\cite{Yang:2025neu},
and polarized dipole scattering amplitudes for small-$x$ helicity evolution~\cite{Dumitru:2024pcv}.
Our numerical results for the helicity flip two-gluon exchange
provide a first idea about its potential magnitude at moderate $x$,
and they could be used in the future as initial conditions for QCD
BFKL evolution to small $x$~\cite{Hatta:2022bxn,Agrawal:2023mzm}. Furthermore, the expressions below demonstrate explicitly
that the helicity flip two-gluon exchange is indeed eikonal, and they clarify its origin as due to
orbital angular momentum transfer to the partons in the proton. The fact that parton orbital
angular momentum modifies certain amplitudes even qualitatively is well known; for example,
it  changes
the asymptotic behavior of the form factor ratio $Q^2 F_2(Q^2)/F_1(Q^2)$ from constant~\cite{Brodsky:1974vy}
to a logarithmic rise~\cite{Belitsky:2002kj}. In the present context, it gives rise to a non-vanishing two-gluon
exchange amplitude with proton helicity flip.

The eikonal dipole amplitude~(\ref{eq:pom1}) in the two gluon exchange approximation, for charges in the
fundamental representation, is given by \cite{Dumitru:2018vpr}
\bea
\calP_{\Lambda \Lambda'}(\rp, \delp) &=&
\frac{g^4 C_F}{2} \int\frac{\rmd^2 \qp}{(2\pi)^2} 
\frac{\cos\left(\frac{\rp\cdot\delp}{2}\right) - \cos\left(\rp\cdot(\qp-\delp)\right)}{\qp^2 (\qp-\delp)^2}
\,\, G_{2,\Lambda\Lambda'}(\qp,\delp-\qp)~.
\label{eq:P_Lambda-Lambda'}
\eea
{Here, $g^2/4\pi = \alpha_s$ is the QCD coupling; for the figures below we chose $\alpha_s=0.35$.}
$G_{2,\Lambda \Lambda'}(\qop,\qtp)$ represents the matrix
element
of two eikonal color charge operators $J^{+a}(\qp) = \int \rmd x^- J^{+a}(x^-,\qp)$
between proton states with helicities $\Lambda$ and $\Lambda'$:
\be
\langle K,\Lambda'|\, J^{+a}(\qop) J^{+b}(\qtp)\, |P,\Lambda\rangle
 = \frac{1}{2}\delta^{ab}\,\, 16\pi^3\, P^+\, \delta(P^+-K^+)\,
  \delta^{(2)}(\Pp - \Kp - \qop-\qtp)\,\,
  G_{2,\Lambda \Lambda'}(\qop,\qtp)\,,
\ee
given by
\be
\begin{split}
 G_{2,\Lambda \Lambda'}(\qop,\qtp) & =
 \int [\rmd x_i] \int [\rmd^2 k_i] \sum\limits_{\{\lambda_i\}}\sum_j
 \Big[\Phi^*_{\Lambda'}(\lambda_i,x_i,\kpij)\Psi^*(x_i,\kpij)\\
  & -\frac{1}{2}\sum_{j'\ne j}
   \Phi^*_{\Lambda'}(\lambda_i,x_i,\kpijj)\Psi^*(x_i,\kpijj)
   \Big]\Phi_{\Lambda}(\lambda_i,x_i,\kpi)\Psi(x_i,\kpi)\,,
 \label{eq:G2}
\end{split}
\ee
with  $\kpij = \kpi + (x_i-\delta_{ij})(\qop + \qtp)$,
$\kpijj = \kpi + x_i (\qop + \qtp) -\qop
\delta_{ij} -\qtp \delta_{ij'}$. The explicit form of the integration
measures over quark LC momentum fractions $x_i$ and transverse momenta $\kpi$ is
given in Appendix~\ref{sec:LFcq-Model}.
For the spatial wave function $\Psi(x_i,\kpi)$ we employ a simple model due to
Schlumpf~\cite{Schlumpf:1992vq,Brodsky:1994fz}.
{\color{black} The spin flip two-gluon exchange amplitude has already been studied
long ago in a non-relativistic quark-diquark model of the proton~\cite{Zakharov:1989bh,Kopeliovich:1989hp}.
It was found that in the soft regime the spin–flip coupling
is small and that it is quite sensitive to diquark correlations. Here, we restrict
to hard gluon exchanges where $|\qop|, |\qtp| \gg \Lambda_\mathrm{QCD}$ due to the fact
that the $t$-channel gluons also couple to a small dipole. In this regime diquark correlations
in the proton may be less significant although such questions would deserve further scrutiny in the future.
}

The helicity wave functions $\Phi_\Lambda(\lambda_i,x_i,\kpi)$
for a proton with helicity $\Lambda$ are obtained
through a Melosh transformation of rest frame Pauli spinors to the
light
front~\cite{Schlumpf:1992vq,Brodsky:1994fz,Ma:1998ar,Pasquini:2008ax}. 
Their explicit expressions for $\Lambda=+1$ and the $|uud\rangle$ 
flavor state are given in
ref.~\cite{Pasquini:2008ax}, for example. In the collinear limit, the only non-zero
functions for $\Lambda=+1$ are $\Phi_+(++-,x_i,\kpi=0) = 2/\sqrt{6}$,
$\Phi_+(+-+,x_i,\kpi=0) = \Phi_+(-++,x_i,\kpi=0) = -1/\sqrt{6}$, i.e.\ they
reduce to the well known non-relativistic SU(2) spin wave functions, as
the Melosh transformation for $\kpi=0$ is trivial. Note that in this
case the product $\Phi^*_+(\lambda_i,x_i,\kpi=0)\, \Phi_-(\lambda_i,x_i,\kpi=0)
=0$ for any given set $\{\lambda_i\}$ of quark helicities:
a helicity flip of the proton is not possible when the quarks
are collinear and carry no orbital angular momentum. However,
this is no longer the case when $\kpi \neq 0$ due to the fact that the
Melosh transformations depend on the quark transverse momenta.  
From the explicit form of the helicity wave functions $\Phi_\Lambda(\lambda_i,x_i,\kpi)$
one may also verify that $G_{2,\Lambda \Lambda'}(\qop,-\qop) \sim \delta_{\Lambda \Lambda'}$
i.e.\ the helicity flip amplitude vanishes in the forward limit.

In practice, we compute $\calP_{\Lambda\Lambda'}(\rp,\delp)$ numerically from \eqref{eq:P_Lambda-Lambda'} and obtain the scalar functions $\calP$, $\calP_S$ and $\calP^\perp_{S}$ via \eqref{eq:pom2}. For the SIP we extract $\calP_0(r_\perp,\Delta_\perp)$ and $\calP_\epsilon(r_\perp,\Delta_\perp)$ from $\calP_{+1,+1}(\rp,\delp)$.
For $\calP_S(\rp,\delp)$ and $\calP^\perp_S(\rp,\delp)$ we determine the leading Fourier harmonics that appear in the amplitudes \eqref{eq:Msdp} and \eqref{eq:Msdpperp}. A convenient way to extract them is through the following angular projections
\be
\begin{split}
&\calP_{S0}(r_\perp,\Delta_\perp) = 8\int_0^{2\pi} \frac{\rmd \phi_\Delta}{2\pi} \int_0^{2\pi} \frac{\rmd\phi_r}{2\pi} \cos(\phi_\Delta)\sin(2\phi_r) \, {\rm Im}\calP_{-1,+1}(\rp,\delp)\,,\\
& \calP^\perp_{S0}(r_\perp,\Delta_\perp) = 2\int_0^{2\pi} \frac{\rmd \phi_\Delta}{2\pi} \int_0^{2\pi} \frac{\rmd\phi_r}{2\pi} \cos(\phi_\Delta) \, \left[{\rm Re}\calP_{-1,+1}(\rp,\delp) - 2 \sin(2\phi_r){\rm Im}\calP_{-1,+1}(\rp,\delp)\right]\,.
\end{split}
\ee

\begin{figure}[htb]
  \begin{center}
  \includegraphics[scale = 0.7]{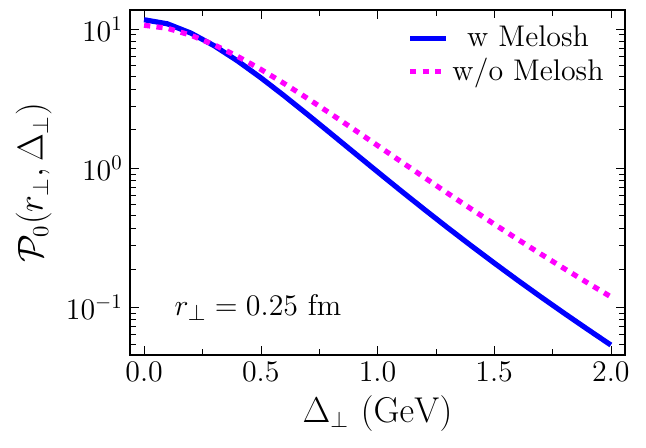}
  \includegraphics[scale = 0.7]{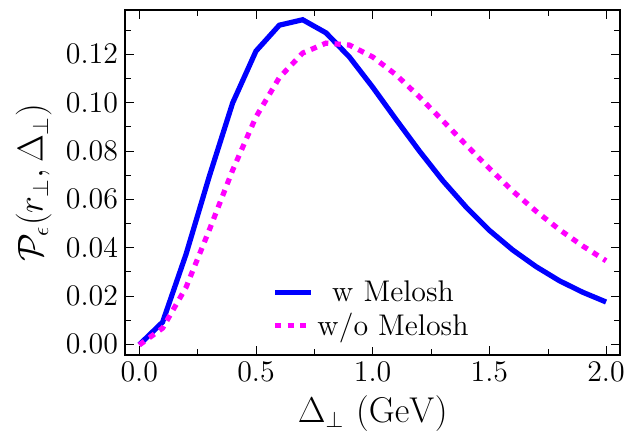}
  \end{center}  \vspace*{-0.5cm}
  \caption{Left: isotropic Pomeron: $\calP_0(r_\perp, \Delta_\perp)$ with $y$-axis on a log scale. Right: elliptic Pomeron $\calP_\epsilon(r_\perp, \Delta_\perp)$.}
  \label{fig:Pomeron2}
\end{figure}
Fig.~\ref{fig:Pomeron2} shows $\calP_0$ (left) and $\calP_\epsilon$ (right) as functions of $\Delta_\perp$ 
for $r_\perp = 0.25$~fm which corresponds approximately to the size of a $J/\psi$ or $\psi(2S)$. 
The Melosh rotation gives a percent-level correction in the forward $\Delta_\perp \to 0$ limit for $\calP_0$. Increasing $\Delta_\perp$ increases the phase space for the orbital motion of quarks and so the effect of the Melosh rotation becomes
more important, leading to corrections of several tens of percent at $\Delta_\perp \sim 1-2$ GeV. 

\begin{figure}[htb]
  \begin{center}
  \includegraphics[scale = 0.75]{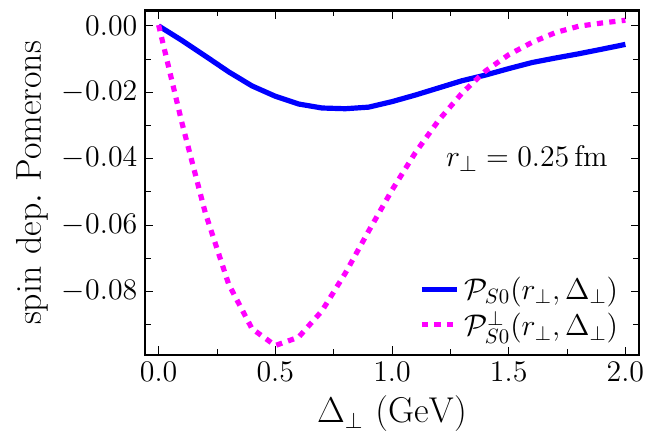}
  \includegraphics[scale = 0.75]{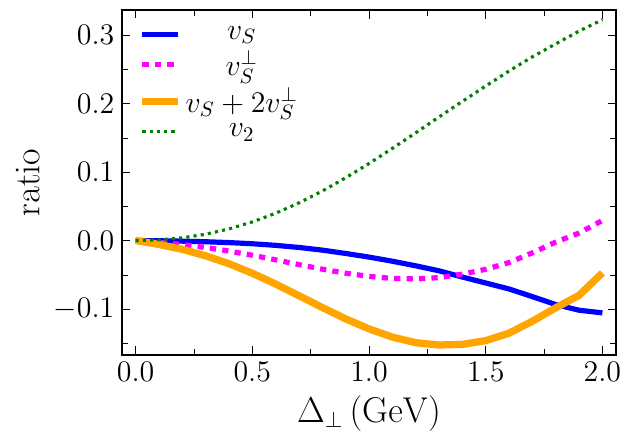}
  \end{center}  \vspace*{-0.5cm}
  \caption{Left: first harmonics of the spin-dependent Pomerons $\calP_{S,0}(r_\perp,\Delta_\perp)$ and $\calP^\perp_{S,0}(r_\perp,\Delta_\perp)$. Right: their ratios to the isotropic Pomeron $\calP_0(r_\perp,\Delta_\perp)$. For comparison we also plot $v_2 = \calP_\epsilon/\calP_0$.}
  \label{fig:ratios}
\end{figure}
In Fig.~\ref{fig:ratios} we show the SDPs $\calP_{S0}$ and $\calP_{S0}^\perp$ and their ratio to the isotropic Pomeron: $v_S = \calP_{S0}/\calP_0$ and $v^\perp_S = \calP^\perp_{S0}/\calP_0$. We also compare to the $v_2$-coefficient defined
from the elliptic Pomeron: $v_2 = \calP_\epsilon/\calP_0$. 
{According to the model computation, both $v_S$ and $v_S^\perp$ are negative and may reach
magnitudes of a few percent as $\Delta_\perp$ increases. The combination $v_S + 2 v_S^\perp$ that enters the helicity flip amplitudes can reach up to $\sim 15\%$ for $\Delta_\perp \sim 1-2$ GeV.}

\section{Summary and Discussion}
\label{sec:conc}
In this paper we have computed the amplitudes for vector meson (VM) production due
to eikonal BFKL (hard) Pomeron exchange, for all helicity configurations in the initial and final states,
including helicity flip of the proton. These amplitudes are important for 
production at high energies and high momentum transfer.

SCHC in exclusive VM production at leading power in energy asserts
that no helicity flip occurs
in the transition of the photon to the $J/\psi$ or $\psi(2S)$ meson, nor in the transition
of the incoming to the scattered proton.
This assertion holds in the forward limit of vanishing momentum transfer $\Delta_\perp$.
Corrections to SCHC arise at non-zero $\Delta_\perp$; we have argued that these originate not
only from higher Fourier harmonics of the usual non-flip BFKL Pomeron but also from
the ``spin dependent Pomeron" associated with a helicity flip of the proton. 

The H1 and ZEUS experiments at
HERA, and the ALICE experiment at the LHC, have tested helicity conservation in the
photon to VM transition, and found it to be satisfied. However, we have argued
that helicity non-conservation in this transition is suppressed not only by powers of
the size of the VM times the momentum transfer but also by, either, powers of the
(small) LC momentum imbalance $z-\bar z$ of quarks in the VM state, or by the (small) ratio
of the ``elliptic" to the isotropic BFKL unintegrated gluon distribution.
On the other hand, helicity flips of the proton are not suppressed by powers of $z-\bar z$
but only by the ratio of the helicity flip vs.\ non-flip unintegrated gluon distributions.
At small $\Delta_\perp \ll m_c$ a collinear limit could be employed~\cite{Frankfurt:1997fj,Ivanov:2004vd} to relate
the ``spin dependent Pomeron" to the GPD $E_g(x,t)$.

According to our numerical estimates the contribution to the
$\gamma + p \to V+p$ cross section from proton helicity flip amplitudes is small when $\Delta_\perp R_V <
1$ but could reach a significant level for large
$\Delta_\perp R_V \sim 1$, where $R_V$ denotes the size of the VM.
{In this regime, the dipole (or $k_\perp$-) factorization approach used here provides
corrections to the collinear limit.} 
A natural observable sensitive to proton helicity flips from spin-dependent Pomerons would be spin asymmetries in exclusive $J/\psi$ production \cite{Koempel:2011rc,Lansberg:2018fsy}, which could be measured
in ultraperipheral $p^\uparrow A$ collisions at RHIC \cite{RHICSPIN:2023zxx} and/or the LHC \cite{Hadjidakis:2018ifr,LHCspin:2025lvj}. 
However, as we have pointed out, single target spin asymmetries vanish in the eikonal limit, i.e.\ they are
suppressed by powers of energy and may be difficult to access experimentally in high-energy collider experiments.
Our proposal is to consider double spin asymmetries in $ep$ collisions with transversely polarized proton and longitudinally polarized electron in the initial state, requiring also the detection of the recoil electron.
We intend to provide numerical estimates for EIC kinematics in the future.
If such measurement can be performed at the EIC then the expressions
we derived will provide a basis for relating data for high momentum transfer
to the dipole formalism.
\\

\begin{acknowledgments}
We thank Y.~Hatta, Yu.~Kovchegov, and K.~Kumeri\v cki for useful discussions, and R.~Venugopalan for stimulating ideas that initiated this work. {\color{black} We thank M.~Spinelli for pointing out an error in 
an earlier version of this paper}. A.~D.\ acknowledges support
by the DOE Office of Nuclear Physics through Grant DE-SC0002307, and the hospitality of the EIC Theory Institute at Brookhaven National Laboratory from Sept.\ 2024 to Jan.\ 2025 where most of this work was performed. S. B. is supported by the Croatian Science Foundation (HRZZ) no. 5332 (UIP-2019-04).
\end{acknowledgments}

\bibliography{references}

\appendix

\section{Photon-vector meson overlaps}
\label{sec:A_i(r,z)}

The light-cone wave function $\gamma-V$ overlaps can be written in the
form~\cite{Mantysaari:2020lhf}
\be
\calA_{\lambda\lambdab}(\rp,\delp) = e q_c \int_z \rme^{-\rmi \dlp\cdot \rp} \int\frac{\rmd^2 \lp}{(2\pi)^2} \frac{\rme^{i\lp\cdot\rp}}{\lp^2 + \varepsilon^2} \int\frac{\rmd^2 \lp'}{(2\pi)^2}\rme^{-\rmi \lonp\cdot \rp} \phi(\lonp,z) \frac{1}{z\bar{z}} A_{\lambda \lambdab}(\lp,\lonp,z)\,.
\ee
Here, $q_c = 2/3$ is the fractional electric charge of the $c$-quark, $\phi(\lonp,z)$ is the VM wave function, $\lonp = \lp' - z\delp$, and $\varepsilon^2  = m_c^2 + z\bar{z}Q^2$.
The function $A_{\lambda \lambdab}(\lp,\lonp,z)$ can be expressed as a Dirac trace
\be
A_{\lambda \lambdab}(\lp,\lonp,z) = \frac{1}{(2q^-)^2} {\rm tr}\left[(\slashed{l} + m_c) \slashed{\epsilon}(\lambda,q)(\slashed{l}-\slashed{q} + m_c)\gamma^- (\slashed{l}' - \slashed{\Delta}+m_c)\slashed{E}(\lambdab,\delp)(\slashed{l}' + m_c)\gamma^-\right]\,,
\ee
with $\epsilon^\mu(\lambda,q)$ and $E^\mu(\lambdab,\Delta)$ the polarization of the photon and the VM, respectively. We have $\epsilon^\mu(\lambda = 0,q) = (Q/q^-,0,\boldsymbol{0}_\perp)$ and $\epsilon^\mu(\lambda = \pm 1,q) = (0,0,\epsp^\lambda)$, where $\epsp^\lambda = (-\lambda,-\rmi)/\sqrt{2}$. The VM polarization is
\be
E^\mu(\lambdab = 0, \Delta) = \frac{1}{M_V}\Delta^\mu - \frac{M_V}{\Delta^-} n^\mu\,, \qquad E^\mu(\lambdab = \pm 1,\Delta) = \left(\frac{\epsp^{\lambdab}\cdot\delp}{\Delta^-},0,\epsp^{\lambdab}\right)\,,
\ee
where $M_V$ is the VM mass. $l$ and $l'$ represent on-shell momenta with $l^- = l'^- = z q^-$.
The traces $A_{\lambda \lambdab}(\lp,\lonp,z)$ evaluate to
\be
\begin{split}
& A_{\lambda = \pm 1, \lambdab = 0}(\lp,\lonp,z) = - 4 M_V {z\bar{z}} (z-\bar{z})(\epsp^\lambda\cdot\lp)\,,\\
& A_{\lambda = \pm 1,\lambdab = \pm 1}(\lp,\lonp,z) = 2 \left[(\epsp^\lambda \cdot \epsp^{\lambdab *})(\lp\cdot \lonp + m^2) + (z-\bar{z})^2 (\epsp^\lambda \cdot \lp)(\epsp^{\lambdab *} \cdot \lonp) - (\epsp^\lambda \cdot \lonp)(\epsp^{\lambdab *} \cdot \lp)\right]\,,
\end{split}
\label{eq:traces1}
\ee
From these we can work out the overlaps $\calA_{\lambda\lambdab}(\rp,\delp)$ to uncover the decomposition in \eqref{eq:overlaps} and the associated functions $\calA_{TL}(r_\perp,z)$ and $\calA_{TT}^{nf,f}(r_\perp,z)$.
For $TT$ polarizations, for example, after performing the integrals over $\lp$ and $\lp'$ we obtain
\be
\begin{split}
\calA_{\lambda = \pm 1\lambdab = \pm 1}(\rp,\delp) = \frac{eq_c}{\pi}\int_z\rme^{-\rmi \dlp \cdot \rp}\frac{1}{z\bar{z}} & \bigg[(\epsp^\lambda \cdot \epsp^{\lambdab *})
\left(-\varepsilon K_1(\varepsilon r_\perp)\frac{\pd \phi_T}{\pd r_\perp} + m_c^2 K_0(\varepsilon r_\perp) \phi_T(r_\perp,z)\right)\\
& + (\epsp^\lambda \cdot \hat{\boldsymbol{r}}_\perp)(\epsp^{\lambdab *} \cdot \hat{\boldsymbol{r}}_\perp) 4 z\bar{z}
\varepsilon K_1(\varepsilon r_\perp) \frac{\pd \phi_T}{\pd r_\perp}\bigg]\,.
\end{split}
\ee
Plugging in the explicit expressions for the transverse polarization vectors we finally arrive at
\be
\begin{split}
&\calA_{TT}^{nf}(r_\perp,z) = \frac{1}{\pi}\frac{1}{z\bar{z}}\left[-(z^2  + \bar{z}^2)
\varepsilon K_1(\varepsilon r_\perp)\frac{\pd \phi_T}{\pd r_\perp} + 
m_c^2 K_0(\varepsilon r_\perp) \phi_T(r_\perp,z)\right]\,,\\
&\calA_{TT}^{f}(r_\perp,z) = \frac{2}{\pi} \varepsilon K_1(\varepsilon r_\perp)\frac{\pd \phi_T}{\pd r_\perp}\,,
\end{split}
\label{eq:ATT}
\ee
that were introduced in the second line of \eqref{eq:overlaps}.
A similar calculation leads to $A_{\lambda 0}(\lp,\lonp,z) = - 4M_V z\bar{z}(z-\bar{z})(\epsp^\lambda \cdot\lp)$ leading to the first line of \eqref{eq:overlaps} and $\calA_{TL}(r_\perp,z)$ given as
\be
\calA_{TL}(r_\perp,z) = \frac{\sqrt{2}\rmi}{\pi} M_V (z-\bar{z}) \varepsilon K_1(\varepsilon r_\perp)
\phi_L(r_\perp,z)\,.
\label{eq:ATL2}
\ee

The functions $\phi_{T,L}(r_\perp,z)$ correspond to the two non-perturbative wavefunctions of the transversely or longitudinally polarized VM, see e.g.\ ref.~\cite{Kowalski:2006hc}.

For the longitudinal photon with virtuality $Q^2$ we find
\be
\begin{split}
& A_{\lambda = 0,\lambdab = 0}(\lp,\lonp,z) = -8Q M_Vz^2 \bar{z}^2\,,\\
& A_{\lambda = 0,\lambdab =\pm 1}(\lp,\lonp,z) = 4 Q z\bar{z} (z - \bar{z}) (\epsp^{\lambdab *} \cdot\lonp)\,.
\end{split}
\ee
The $\gamma^*_L-V$ wave function overlaps are
\be
\begin{split}
& \calA_{\lambda = 0,\lambdab = 0}(\rp,\delp) = e q_c \int_z \rme^{-\rmi \dlp\cdot\rp} \calA_{LL}(r_\perp,z)\,,\\
& \calA_{\lambda = 0,\lambdab = \pm 1}(\rp,\delp) = e q_c \lambdab \rme^{-\rmi \lambdab\phi_r} \int_z \rme^{-\rmi \dlp\cdot\rp} \calA_{LT}(r_\perp,z)\,,
\end{split}
\label{eq:overlapsL}
\ee
where
\be
\begin{split}
\calA_{LL}(r_\perp,z) = -\frac{4}{\pi}Q M_V z\bar{z} K_0(\varepsilon r_\perp) \phi_L(r_\perp ,z)\,,\\
\calA_{LT}(r_\perp,z) = -\frac{\sqrt{2}\rmi}{\pi} Q (z-\bar{z}) K_0(\varepsilon r_\perp) \frac{\partial\phi_T}{\partial r_\perp}\,.
\end{split}
\label{eq:Along}
\ee


\section{Pomeron amplitude at moderately small $x$ from a light-front
constituent quark model}
\label{sec:LFcq-Model}

We now provide explicit expressions for the Pomeron exchange
amplitudes at moderately small $x$ off a proton described by a lightfront
quark model. Our expressions generalize similar expressions from
ref.~\cite{Dumitru:2018vpr,Dumitru:2020fdh} where, however, a helicity
flip of the proton had not been considered.

We write the proton state in the LF quark model in the form
\be
\begin{split}
  |P,\Lambda\rangle &= \int[\rmd x_i]\int\left[\rmd^2\kpi\right]
  \sum_{j_1, j_2, j_3} \frac{\epsilon^{j_1 j_2 j_3}}{\sqrt{N_c!}}
  \, \sum_{\{\lambda_i\}} \Phi_\Lambda(\lambda_i,\kpi)\,
  \Psi(x_i,\kpi)\,\, |\{x_i P^+,\kpi,\lambda_i,j_i\}\rangle\,.
\end{split}
\label{eq:|P>}
\ee
Here, $x_i$ denotes the LC momentum fraction of the $i^\mathrm{th}$ quark
where $i=1\dots N_c=3$;
$\kpi$ is its transverse momentum relative to the CM transverse
momentum $\Pp$ of the proton; $j_i$ refers to its color; $\lambda_i$ to its
helicity. The integrations over $x_i$ and $\kpi$ are given by
\be
   [\rmd x_i] = \prod_{i=1\cdots 3} \frac{\rmd x_i}{2x_i}\, \delta
   \left(1 - \sum_i x_i\right)\ ,\ \
        \left[\rmd ^2 \kpi\right] = \prod_{i=1\cdots 3}
\frac{\rmd ^2 \kpi}{(2\pi)^3}\, (2\pi)^3\,\delta^{(2)} \left( \sum_i \kpi
\right)~.
\ee
The spatial wave function $\Psi(x_i,\kpi)$ is symmetric under the
exchange of any two quarks and invariant under simultaneous rotations
or reflections of all transverse momenta $\kpi$. For numerical
estimates we employ a simple model due to
Schlumpf~\cite{Schlumpf:1992vq,Brodsky:1994fz},
\be
\Psi(x_i,\kb_i) \sim \sqrt{x_1 x_2 x_3}\,\, e^{-\mathcal{M}^2/2\beta^2}~~,
~~~~~
\mathcal{M}^2 = \sum_i \frac{\kpi^2+m_q^2}{x_i}~.
\ee
The quark mass $m_q=0.26$~GeV and the parameter $\beta=0.55$~GeV have
been tuned to reproduce the electromagnetic ``radius'' and the anomalous magnetic
moments of proton and neutron\footnote{We have omitted the flavor wave
function in~(\ref{eq:|P>}) as this does not play a role in our
analysis.}; they also lead to reasonably good agreement of the Dirac
and Pauli form factors of the proton with data. The normalization of
the spatial wave function follows from the requirement that
$\left<K,\Lambda' | P,\Lambda\right> = 16 \pi^3 P^+\delta(P^+ - K^+)
\delta^{(2)}(\Pp - \Kp) \delta_{\Lambda\Lambda'}$.

The helicity wave functions $\Phi_\Lambda(\lambda_i,\kpi)$
in~(\ref{eq:|P>}) for a proton with helicity $\Lambda$ are obtained
through a Melosh transformation of rest frame Pauli spinors to the
LF. Their explicit expressions for $\Lambda=+1$ are given in
ref.~\cite{Pasquini:2008ax}; here we add only that the functions for
$\Lambda=-1$ are obtained by a sign flip of $\Phi$, and the exchange
$k_{iL} \leftrightarrow - k_{iR}$, where $k_{R/L}=k_\perp^1 \pm \rmi k_\perp^2 = k_\perp e^{\pm \rmi \phi_k}$, and
$\lambda_i \to - \lambda_i$.\\

The eikonal dipole amplitude at order $(g A^+)^2$ (two gluon exchange)
has been expressed in eq.~(\ref{eq:P_Lambda-Lambda'}) in terms of
the color charge correlator $G_{2,\Lambda \Lambda'}(\qop,\qtp)$
of eq.~(\ref{eq:G2}). This color charge
correlator vanishes, for any $\Lambda, \Lambda'$ when either $q_{1\perp}$
or $q_{2\perp}$ go to zero as gluons with wavelength greater than the size
of the proton decouple. Furthermore, in the forward $t\to0$ limit,
$G_{2,\Lambda \Lambda'}(\qp,-\qp) \sim
\delta_{\Lambda\Lambda'}$, i.e.\ a helicity flip of the proton is then
not possible\footnote{This statement refers specifically to two gluon exchange where in the
$t\to0$ limit $\qonp = - \qtwp$ are anti-collinear. For $C$-odd three gluon exchange a helicity flip of
the proton {\em is} possible even in the forward limit where it is related to the gluon Sivers
function~\cite{Zhou:2013gsa,Boussarie:2019vmk,Kovchegov:2021iyc,Benic:2024fbf}.}.
Eq.~(\ref{eq:G2}) provides an explicit illustration for
the origin of the proton helicity flip in the eikonal limit: it is due
to the Melosh transformation from Pauli to LC helicity spinors which
depends on the quark transverse momenta~\cite{Ma:1998ar,Pasquini:2008ax}. A transverse momentum
transfer can then lead to non-zero overlap of the helicity wave
functions for $\Lambda \ne \Lambda'$ unlike in a non-relativistic
quark model of the proton.

\section{GTMDs of the eikonal dipole and the spin dependent Pomerons}
\label{sec:gtmds}
The connection of the Pomeron $\calP_{\Lambda\Lambda'}(\rp,\delp)$ to the GTMDs has been established in 
refs.~\cite{Boussarie:2019vmk,Hagiwara:2020mqb}. From eqs.~(14) and (15) in \cite{Hagiwara:2020mqb} we have
\be
\begin{split}
\int \rmd^2 \rp \rme^{-\rmi \kp\cdot\rp} \calP_{\Lambda\Lambda'}(\rp,\delp) & = - \frac{g^2 (2\pi)^3}{4 N_c M (\kp^2 - \delp^2/4)}\Bigg[M \delta_{\Lambda\Lambda'} f_{1,1}(\kp,\delp) + \Delta_\perp \frac{\kp^2}{M^2} \delta_{\Lambda,-\Lambda'} \Lambda\rme^{\rmi \Lambda \phi_k}f_{1,2}(\kp,\delp)\\
& + \Delta_\perp \Lambda \delta_{\Lambda,-\Lambda'} \rme^{\rmi \Lambda \phi_\Delta}\left(f_{1,3}(\kp,\delp) - \frac{1}{2}f_{1,1}(\kp,\delp)\right)\Bigg]\,,
\end{split}
\ee
where the $f_{1,i}(\kp,\delp)$ are the real parts of the GTMDs, and $M$ is the proton mass. The imaginary parts of the
GTMDs are related to the Odderons which are not relevant in the context of VM production. The Pomeron amplitudes are related to the $f_{1,i}$ as follows~\cite{Hagiwara:2020mqb}
\be
\begin{split}
\calP(\rp,\delp) &= \frac{g^2 (2\pi)^3}{4 N_c} \int \frac{\rmd^2 \kp}{(2\pi)^2} \rme^{\rmi \kp\cdot\rp}\frac{f_{1,1}(\kp,\delp)}{\kp^2 - \delp^2 /4}\,,\\
\calP_S(\rp,\delp) &= \frac{g^2 (2\pi)^3}{4 N_c}\frac{\Delta_\perp}{M} \int \frac{\rmd^2 \kp}{(2\pi)^2} \rme^{\rmi \kp\cdot\rp}\left[2\frac{(\rp\cdot\kp)^2}{\rp^2 M^2} - \frac{\kp^2}{M^2}\right]\frac{f_{1,2}(\kp,\delp)}{\kp^2 - \delp^2 /4}\,,\\
\calP^\perp_S(\rp,\delp) &= \frac{g^2 (2\pi)^3}{4 N_c}\frac{\Delta_\perp}{M} \int \frac{\rmd^2 \kp}{(2\pi)^2} \rme^{\rmi \kp\cdot\rp}\Bigg[-\frac{1}{2} f_{1,1}(\kp,\delp) + \left(-\frac{(\rp\cdot\kp)^2}{\rp^2 M^2} + \frac{\kp^2}{M^2}\right) f_{1,2}(\kp,\delp)\\
& ~~~~~~~~~~~~~~~~~ + f_{1,3}(\kp,\delp)\Bigg]\frac{1}{\kp^2 - \delp^2 /4}\,.
\end{split}
\ee

\end{document}